\documentclass[a4paper,11pt]{article}

\pdfoutput=1
\usepackage{jcappub}
\usepackage{graphicx,subfig}
\usepackage{amssymb,amsmath,bm}
\usepackage[usenames,dvipsnames]{xcolor}
\usepackage{hyperref}
\usepackage{multirow}

\usepackage[T1]{fontenc}

\newcommand{\lya}{Lyman-$\alpha$}
\newcommand{\vk}{\mathbf{k}}

\newcommand{\bdla}{b_{\rm DLA}}
\newcommand{\hMs}{\ h^{-1} M_\odot}
\newcommand{\nv}{\hat{\bf n}}

\title{Bias of Damped Lyman-$\alpha$ systems from their cross-correlation with CMB lensing}

\author[a,1]{D. Alonso,\note{Author list is alphabetized}}
\author[b,c]{J. Colosimo,}
\author[d]{A. Font-Ribera}
\author[b]{and A. Slosar}

\affiliation[a]{Oxford Astrophysics, Keble Road, Oxford OX1 3RH, UK}
\affiliation[b]{Physics Department, Brookhaven National Laboratory, Upton NY 11973, USA}
\affiliation[c]{School of Physics, Georgia Tech, 837 State St NW, Atlanta, GA 30332, USA}
\affiliation[d]{Department of Physics and Astronomy, University College London, 132 Hampstead Road, London, NW1 2PS, UK}

\emailAdd{david.alonso@physics.ox.ac.uk}
\emailAdd{jcolosimo@gatech.edu}
\emailAdd{a.font@ucl.ac.uk}
\emailAdd{anze@bnl.gov}

\abstract{We cross-correlate the positions of damped \lya\ systems (DLAs) and their parent quasar catalog with a convergence map derived from the Planck cosmic microwave background (CMB) temperature data. We make consistent measurements of the lensing signal of both samples in both Fourier and configuration space. By interpreting the excess signal present in the DLA catalog with respect to the parent quasar catalog as caused by the large scale structure traced by DLAs, we are able to infer the bias of these objects:  $\bdla=2.6\pm0.9$. These results are consistent with previous measurements made in cross-correlation with the \lya\ forest, although the current noise in the lensing data and the low number density of DLAs limits the constraining power of this measurement. We discuss the robustness of the analysis with respect to a number different systematic effects and forecast prospects of carrying out this measurement with data from future experiments.}

\begin{document}
\maketitle
\flushbottom

  \section{Introduction}\label{sec:intro}
    Damped \lya\ systems (DLAs) are absorption features in the spectra of high redshift quasars, with a neutral hydrogen column density $N_{\rm HI} > 2 \times 10^{20} \rm{cm}^{-2}$ \cite{1986ApJS...61..249W}, with damped wings covering a large fraction of the spectra. DLAs are an important source of contamination in studies of the clustering of the \lya\ forest \cite{2006ApJS..163...80M,2012JCAP...07..028F,2017arXiv171106275R,2017arXiv170608532R}, and it is important to model their clustering properties in order to use the \lya\ forest as a cosmological tool. 

    On the other hand, DLAs are very interesting astrophysical objects themselves, since they contain large amounts of neutral hydrogen and are thought to play a key role in the formation of galaxies at high redshift. They are also considered to be the main source of emission in ongoing and future 21cm surveys \cite{2015aska.confE..17A,2014SPIE.9145E..22B,Newburgh2016:1607.02059v1}, where most of the signal is expected to come from galaxies with a high density of neutral hydrogen. Understanding the distribution of neutral hydrogen as traced by DLAs will allow us to make better predictions for the amplitude of the 21cm signal, which is currently uncertain by a factor of a few \cite{2012ApJ...750...38M,2015aska.confE..21S,2017MNRAS.471.1788C}.

    Unfortunately, the low number density of DLAs prevents a direct measurement of their bias from the amplitude of their auto-correlation. However, a first measurement of the clustering strength of DLAs was presented in \cite{2012JCAP...11..059F}. For this, the authors used quasar spectra from the ninth data release (DR9, \cite{2012ApJS..203...21A}) of the Sloan Digital Sky Survey (SDSS-III, \cite{2011AJ....142...72E}), in particular from the Baryon Oscillation Spectroscopic Survey (BOSS, \cite{2013AJ....145...10D}), to measure at high significance the cross-correlation of DLAs with the \lya\ forest absorption. Assuming a cosmological model, as well as bias parameters for the \lya\ forest, they presented an estimate of the DLA bias of $\bdla = 2.17 \pm 0.20$. Recently, this result has been confirmed by \cite{2018MNRAS.473.3019P}, who used a similar technique on the twelfth data release (DR12, \cite{2015ApJS..219...12A}) of SDSS-III to present an updated measurement of $\bdla = 1.99 \pm 0.11$.

    These results came as a surprise, since they imply that DLAs are hosted by rare and massive halos, with typical masses upwards of $\sim 10^{11} \hMs$. On the other hand, most hydrodynamical simulations predict that DLAs should have a lower clustering strength, since the main contribution to the total HI density in simulations comes from halos of smaller mass \cite{1997ApJ...484...31G,2008MNRAS.390.1349P}, except when strong galactic winds are included \cite{2007ApJ...660..945N,2009MNRAS.397..411T,2012ApJ...761...54E,2014MNRAS.445.2313B}. Resolving these discrepancies might have significant repercussions for the astrophysics of small halos.

    In this paper we set out to measure the large-scale bias of DLAs with an alternative technique, by measuring their cross-correlation with CMB lensing data. Compared to cross-correlation with the other tracers in their vicinity, such as quasars, or the \lya\ forest, this has an advantage that there is no bias factor associated with the CMB lensing map and that systematic effects should be very different from other methods of measuring the DLA bias. As we will discuss later, current data allows a measurement that is consistent with existing determinations, but with large error bars. However, we argue that future data will allow a decisive measurement to be made using this technique.

    In Section \ref{sec:theory} we describe the measurements and outline the relevant calculations for making  theoretical predictions. In Section \ref{sec:data} we discuss the datasets used in the analysis and in Section \ref{sec:results} we present our results. We forecast the ability of future experiments to improve upon this measurements in Section \ref{sec:forecast} and we summarize our findings in Section \ref{sec:conclusions}.

    \begin{figure}
      \centering
      \includegraphics[width=0.7\linewidth]{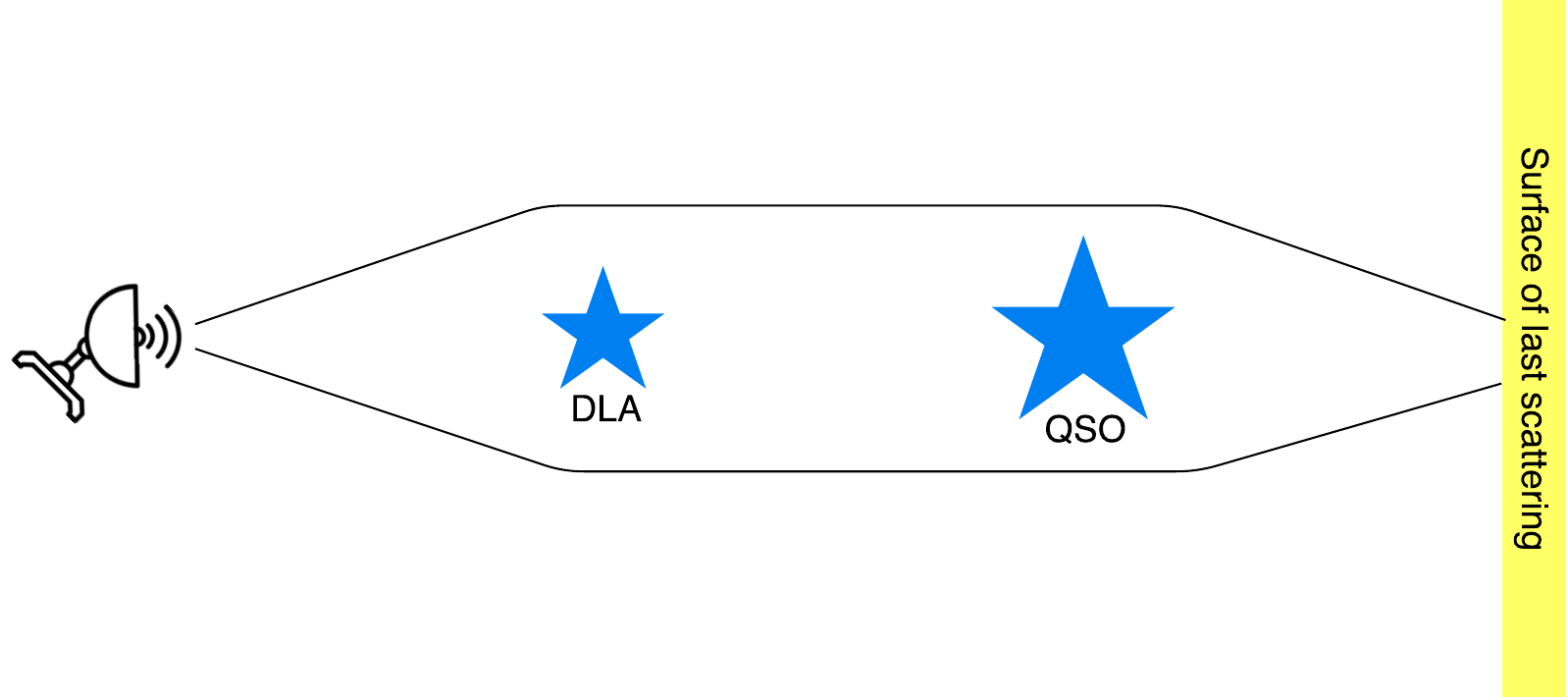}
      \caption{Figure outlining the effect. Photons emitted at the surface of last scattering are deflected by the intervening mass that correlates with positions of DLAs. However, since DLAs are found in quasar spectra, there is a quasar behind every DLA that also lenses the CMB photons. The latter contribution must be accounted for when measuring DLA bias.}\label{fig:scheme}
    \end{figure}

  \section{Theory}\label{sec:theory}
    In Figure \ref{fig:scheme} we show schematically the measurement
    that we are trying to perform. DLAs trace the overall structure of
    the Universe and are linearly biased with respect to the dark
    matter fluctuations $\delta_{\rm DM}$, leading to the following
    Fourier space relation valid on large scales (small wavenumbers $\vk$):
    \begin{equation}
      \bdla (\vk) = b_{\rm DLA} \delta_{\rm DM} (\vk).
    \end{equation}
    Since matter over-densities gravitationally lense the CMB photons, we expect projected DLA overdensities to do the same. Thus, for fixed cosmological parameters, we should be able to infer the bias parameter of the DLAs from the amplitude of the cross-correlation with a CMB lensing map. The linear biasing becomes exact in the large-scale limit, $\vk\rightarrow 0$, and should be a very good approximation over all scales on which the data used in this paper is sensitive.

    A subtlety in this measurement is the fact that DLAs are found in the spectra of distant quasars, and hence there is a quasar behind every DLA. These quasars in turn trace structures at higher redshift behind the DLAs that cause their own lensing of the CMB photons. Since the path-length of the forest is very large, these structures are, to a very good approximation, unrelated. I.e. the quasars are  backlights that illuminate the DLAs in front of them, but are sufficiently far away that they can be considered otherwise uncorrelated with the DLAs. Of course, the only way to subtract the effect of the quasar sample is to subtract an otherwise statistically equivalent sample of quasars but without DLAs in them. As we will later argue, it is sufficient to reweight the quasars based on their redshifts in order to obtain a sample with the same redshift distribution as that subsample which hosts DLAs.

    We will now put this on a firmer mathematical footing. The paths of photons emitted by a source at comoving radial distance $\chi_s$ are perturbed by the intervening matter inhomogeneities through the gravitational lensing effect. In a flat Universe, the angular deflection in the trajectories of these photons arriving from the direction $\nv$ can be written as the gradient of an underlying lensing potential $\Phi_L$, given by \cite{2001PhR...340..291B}:
    \begin{equation}
      \Phi_L(\nv)=\int_0^{\chi_s}d\chi\,\frac{\chi_s-\chi}{\chi_s\,\chi}\,\Phi(\chi\nv,\chi),
    \end{equation}
    where $\Phi(\chi\nv,\chi)$ is the Weyl potential evaluated in the lightcone. The Weyl potential is $\Phi=\psi+\phi$, where $\phi$ and $\psi$ are the two metric potentials in the Newtonian gauge. In the case of CMB lensing, $\chi_s$ is the distance to the surface of last scattering. Finally, the convergence $\kappa$ is given by the Laplacian of $\Phi_L$ on the sphere $\kappa\equiv\nabla_{\nv}^2\Phi_L/2$.

    On the other hand, assuming that DLAs and quasars are biased tracers of the underlying matter density perturbations, their angular overdensity is given by:
    \begin{equation}
      \delta^\alpha(\nv)=\int d\chi\,b_\alpha(\chi)\,p_\alpha(\chi)\,\delta_M(\chi\nv,\chi),
    \end{equation}
    where $\alpha$ denotes either DLAs or QSOs, $b_\alpha$ is the linear bias factor of tracer $\alpha$, $p_\alpha(\chi)$ is its normalized selection function and $\delta_M(\chi\nv,\chi)$ is the matter overdensity field in the lightcone. The equation above  assumes that a scale-independent, linear bias factor is enough to characterize the relation between the DLA/QSO distribution and the density field, which is known to be a good approximation on large, linear scales. The noise properties of the Planck lensing map, as well as the high shot noise of both source catalogs guarantee that only linear scales ($r\gtrsim20\,{\rm Mpc}/h$) contribute significantly to the total signal-to-noise of this measurement, and therefore this approximation should suffice.

    The angular cross-power spectrum between $\kappa$ and $\delta^\alpha$, defined as the two-point correlator of the harmonic coefficients of both quantities, $C^{\kappa\alpha}_\ell\equiv\langle\kappa_{\ell m}\delta^\alpha_{\ell m}\rangle$, can be related to the matter power spectrum $P(k,z,z')$ as:
    \begin{equation}
      C^{\kappa\alpha}_\ell=\frac{2}{\pi}\int dk\,k^2 \int d\chi \int d\chi' P(k,z(\chi),z(\chi')) \, W^\alpha_\ell(\chi,k) \, W^\kappa_\ell(\chi',k),
    \end{equation}
    where
    \begin{equation}
      W^\alpha_\ell(\chi,k)=p_\alpha(\chi)b_\alpha(\chi)j_\ell(k\chi),\hspace{12pt}
      W^\kappa_\ell(\chi,k)=\frac{3}{2}H_0^2\Omega_M\frac{\ell(\ell+1)}{k^2\,a(\chi)}\Theta(\chi;0,\chi_s)\frac{\chi_s-\chi}{\chi\,\chi_s}j_\ell(k\chi).
    \end{equation}    
    Here $j_\ell(x)$ is the spherical Bessel function of order $\ell$, $\Theta(x;x_1,x_2)$ is a top-hat window function in the interval $(x_1,x_2)$, and $a(\chi)$ is the scale factor at which the radial comoving distance is $\chi$. We will also account for the effect of lensing magnification in the angular distribution of quasars. This can be easily done by modifying $W^\alpha_\ell(\chi,k)$:
    \begin{equation}
      W^\alpha_\ell(\chi,k)=\left[p_\alpha b_\alpha(\chi)+(5s-2)\frac{3H_0^2\Omega_M}{2k^2a(\chi)}\int_\chi^{\chi_{\rm H}}d\chi'\,p_\alpha(\chi)\frac{\chi'-\chi}{\chi\chi'}\right]j_\ell(k\chi),
    \end{equation}
    where $\chi_{\rm H}$ is the distance to the horizon, and we fix the slope of the quasar magnitude distribution to $s=0.2$ \cite{2005ApJ...633..589S}.

    To simplify the calculation of this observable we use Limber's approximation, which is equivalent to substituting $j_\ell(x)\sim\sqrt{\pi/2}\,\delta(\ell+1/2-x)/\sqrt{\ell+1/2}$. This approximation is excellent for the wide radial kernels of the three tracers explored here. The expression for the angular power spectrum in this approximation becomes much simpler:
    \begin{equation}
      C^{\kappa\alpha}_\ell=b_\alpha\,\frac{3}{2}H_0^2\Omega_M\frac{\ell(\ell+1)}{(\ell+1/2)^2}\int dk\,k\,\frac{p_\alpha(\chi_\ell)}{a(\chi_\ell)}\Theta(\chi_\ell;0,\chi_s)\frac{\chi_s-\chi_\ell}{\chi_s}P(k,\chi_\ell).
    \end{equation}
    Here we have defined $\chi_\ell\equiv(\ell+1/2)/k$. We have also assumed that the DLA/QSO bias is redshift-independent. While it would be very interesting to study the redshift dependence of this quantity, the noise in the current data does not allow for a detailed study beyond the overall normalization of their clustering amplitude.

    Finally, we will also be interested in computing the two-point correlation function between $\kappa$ and $\delta_\alpha$ in configuration space. This quantity is defined as $\xi^{\kappa\alpha}(\theta)\equiv\langle\kappa(\nv_1)\delta_\alpha(\nv_2)\rangle$, where $\theta$ is the angle between $\nv_1$ and $\nv_2$. In the small-angle limit, it is easy to prove that $C_\ell$ and $\xi(\theta)$ are related by a Hankel transform:
    \begin{equation}
      \xi^{\kappa\alpha}(\theta)=\frac{1}{2\pi}\int d\ell\,\ell\,C_\ell^{\kappa\alpha}\,J_0(\ell\theta),
    \end{equation}
    where $J_0(x)$ is the cylindrical Bessel function of order 0. At this point it is worth noting that our configuration-space analysis, described in Section \ref{sec:results}, uses a filtered version of the convergence map. This can be easily taken into account by modifying the previous equation, substituting $C^{\kappa\alpha}_\ell\rightarrow W^\kappa_\ell\,C^{\kappa\alpha}_\ell$, where $W^\kappa_\ell$ is the filter applied to the convergence map in harmonic space.

    All angular power spectra and correlation functions were computed using the Core Cosmology Library\footnote{CCL is a public library to compute theoretical predictions for multiple cosmological observables. The library is maintained by the Dark Energy Science Collaboration of LSST, and can be found in \url{https://github.com/LSSTDESC/CCL}}, using {\tt CLASS} \cite{2011JCAP...07..034B} to compute the matter power spectrum. In this analysis we fixed all cosmological parameters to their best-fit values found by \cite{2016A&A...594A..13P}.

  \section{Data}\label{sec:data}
    We use the CMB lensing convergence data provided by the Planck Collaboration \cite{2016A&A...594A..15P}. We transformed the provided multipole coefficients $\kappa_{\ell m}$ into a HEALPix map \cite{2005ApJ...622..759G} with resolution $N_{\rm side}=2048$, and verified that this choice of resolution had a negligible impact on our results by re-running the analysis pipeline described in Section \ref{sec:results} for $N_{\rm side}=1024$. When estimating both correlation functions and power spectra we mask this convergence map using the mask provided by Planck, covering around 67\% of the sky. This mask was designed to remove regions dominated by galactic foregrounds as well as strong Sunyaev-Zel'dovich sources. Since these sources lie mostly on redshifts $z\lesssim0.4$, this mask is completely uncorrelated with the quasar and DLA catalogs used here. We note that any residual contamination from extra-galactic foregrounds in the CMB lensing map (mainly the thermal Sunyaev-Zel'dovich effect and the cosmic infrared background) is likely to correlate with the distribution of quasars. We assume this effect to be subdominant, given the statistical uncertainties in our measurement.

    \begin{figure}
      \centering
      \includegraphics[width=0.49\textwidth]{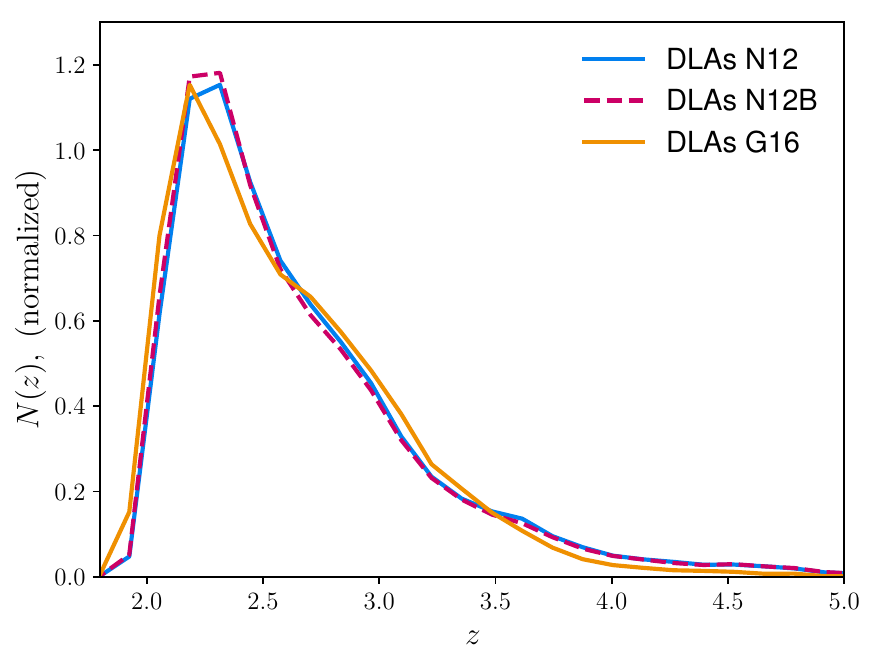}
      \includegraphics[width=0.49\textwidth]{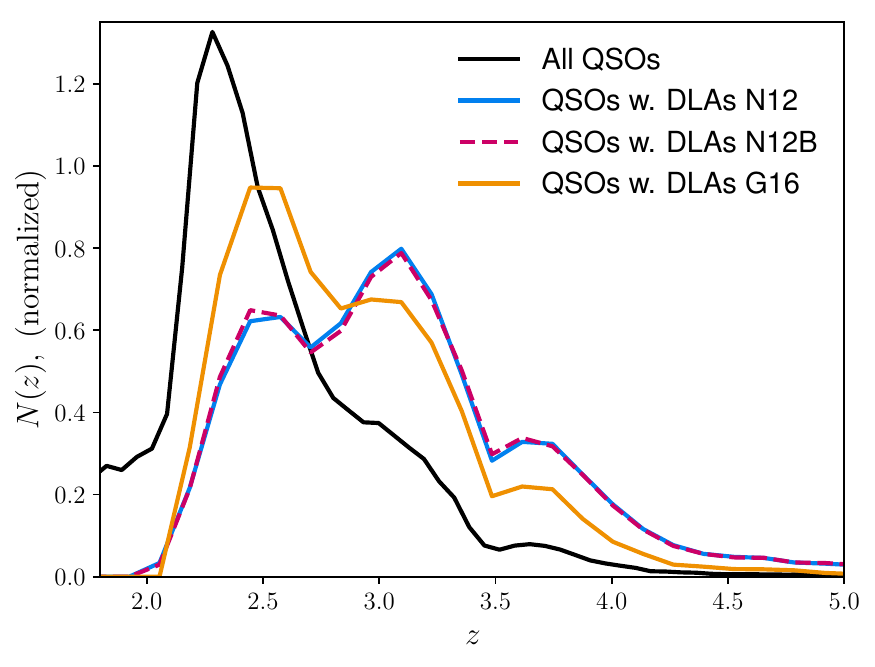}
      \caption{{\sl Left:} redshift distribution of DLAs in the N12 (solid cyan), N12B (dashed dark magenta) and G16 (solid light orange) catalogs. {\sl Right:} redshift distribution of the full DR12 quasar catalog (solid black) as well as that of the QSOs containing DLAs in their spectra in the N12 (solid cyan), N12B (dashed dark magenta) and G16 (solid light orange) DLA catalogs. All distributions are normalized to unit area.}\label{fig:nz}
    \end{figure}

    \begin{figure}
      \centering
      \includegraphics[width=0.55\textwidth]{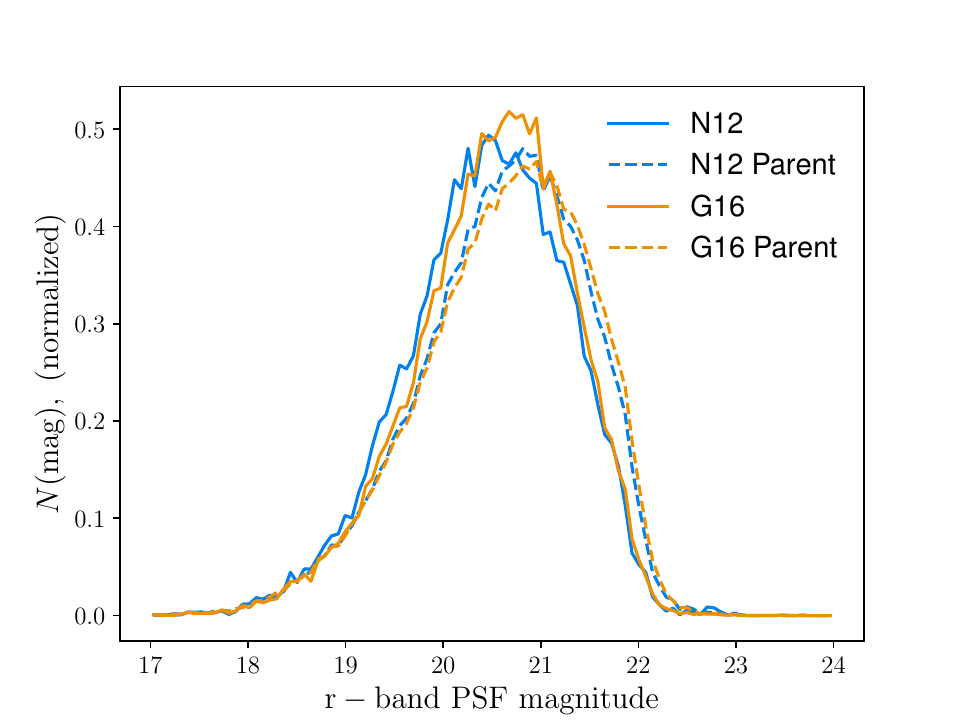}
      \caption{Distribution of $r$-magnitude for quasars where DLAs
        have been found (solid lines) and the parent catalogs (created
      by reweighting the full quasar catalog to match redshift
      distributions, dashed lines) for the  N12 (red) and G16 (blue) catalogs.}\label{fig:umagdist}
    \end{figure}

    We use two different DLA catalogs, both created from quasar spectra from the 12th data release of the Sloan Digital Sky Survey-III (SDSS-III DR12 \cite{2015ApJS..219...12A}). The first catalog, which we will refer to as N12, identified DLAs using an algorithm first described in \cite{2012A&A...547L...1N}, and has been used in the DR12 \lya\ analyses of BOSS (\cite{2017A&A...603A..12B,2017A&A...608A.130D,2018MNRAS.473.3019P}). In short, DLAs are identified by correlating the spectra with synthetic absorption profiles of increasing column density. The total catalog consists of 34,050 systems, with a redshift distribution shown as a solid cyan line in the left panel of Figure \ref{fig:nz}. In Section \ref{sec:results} we also use a subset of this catalog, which we label N12B, where we exclude those systems identified in the spectra with lower signal to noise. 
    The second catalog, which we will refer to as G16\footnote{The catalog is publicly available at \url{http://tiny.cc/dla_catalog_gp_dr12q_lyb}}, used instead an algorithm presented in \cite{2017MNRAS.472.1850G}. In this case the authors used a Gaussian process to model the quasar emission spectra, and used Bayesian model selection to assign a probability of hosting a DLA to each spectrum. We used G16 with an extra quality cut to ensure the purity of the catalog. In particular, we select only the DLA candidates that have a probability $p_{\rm DLA}>0.9$, thus retaining a sample of 35,621 objects. This probability cut-off was chosen to generate a catalog of similar size to the N12 catalog. Since the distribution of probabilities is very bi-modal (most cluster around $p=0$ or $p=1$), the number of DLAs is surprisingly insensitive to the precise choice of probability cut.    
    Note that during the training of the machine learning algorithm used in G16 the authors used information from the DLA catalog of \cite{2012A&A...547L...1N} that used the same algorithm as N12, and therefore we expect the two catalog to be quite similar.  
        
    For both DLA catalogs, we also generate the parent quasar catalog. This is done by taking the full catalog of quasars where a DLA could have been found and reweighting it so that the redshift distribution of quasars where DLA have been found matches that of the reweighted full catalog. These parent catalogs allow us to measure the ``null'' signal, i.e the signal in absence of lensing by the DLAs.

    The properties of the samples are summarized in Figure \ref{fig:nz}. We see that the redshift distribution of DLAs is very similar for both catalogs and peaks at around $z\sim 2.3$. The shape of this distribution is dominated by the total path length of the \lya\ forest available as a function of redshift. Below $z\sim1.8$, the forest goes into UV and is unobservable from the ground. At redshifts above $z>2.5$, the number density of quasars decreases and they become fainter and thus noiser which both act to lower the effective number of DLAs. The right panel of the same figure shows the redshift distribution of the quasars in which these DLAs have been found. We see that the majority of quasars skew towards lower redshifts, but these have very short forests and thus not very many DLAs are found in them. However, the parent quasar redshift distributions for the N12 and G16 catalogs are surprisingly different given the similarity of the DLA distributions plotted in the left panel. This means, that although both have similar redshift distributions, the two samples actually detect \emph{different} DLAs. 

    The method to generate synthetic parent QSO catalogs described above assumes that quasar redshift is the only property that determines the probability of detecting a DLA. We know that this must be an imperfect assumption at some level since, for instance, it is easier to identify DLAs in the spectra of bright quasars. If the clustering of quasars had a strong dependence on their luminosity, this could bias our result, since the clustering of quasars without DLAs would be on average lower than the clustering of quasars with DLAs in their spectra. In order to study this possibility, Figure \ref{fig:umagdist} shows the $r$-band magnitude distribution for the quasars where DLAs have been found (solid lines) and for their synthetic parent catalogs generated using only redshift-dependent weights (dashed). We see that, although the synthetic catalogs are slightly fainter -- this is expected as DLAs are preferentially identified in brighter quasars that have better signal-to-noise in the \lya\ forest region -- the effect is surprisingly small. This result, together with the independence of the quasar bias on luminosity, verified by different analyses using BOSS-like samples \cite{2013JCAP...05..018F,2015MNRAS.453.2779E,2017JCAP...07..017L} gives us confidence that redshift-dependent weights are enough to mimic the angular clustering properties of the DLA parent sample. In particular, the measurements of \cite{2015MNRAS.453.2779E} show a variation of the quasar bias of less than $10\%$ within the magnitude range $-28.74\le M_i\le-23.78$.
    
  \section{Method \& Results}\label{sec:results}
    \subsection{Differential bias measurement}\label{sec:results.diffm}
      As mentioned in Section \ref{sec:data}, the main complication of the measurement we are attempting here is the fact that the DLA catalogs cannot be directly interpreted as a homogeneous point process: the DLAs all lie along the lines of sight of illuminating quasars, which contribute to the total lensing signal. Thus, a stack of the CMB $\kappa$ map around the positions of DLAs will contain contributions from the lensing signals of both the DLAs and the background quasars. The quasar contribution must therefore be estimated and subtracted in order to get at the pure DLA signal.
                  
      Now, let ${\bf d}_{\rm D+Q}$ be a measurement of the lensing signal on DLAs, including the contribution of the background quasars, and let ${\bf d}_{\rm Q}$ be the same measurement performed on the parent QSO sample. Here ${\bf d}$ stands for a set of measurements of either the correlation function or the power spectrum at $N_d$ values of $\theta$ or $\ell$. There are several ways of combining both data vectors to obtain a measurement of the bias of DLAs, $b_{\rm DLA}$. Here we  explore two possibilites:
      \begin{enumerate}
        \item {\bf Direct subtraction.} As an estimator for the DLA-only signal we directly subtract the measured QSO contribution: ${\bf d}_{\rm D}={\bf d}_{\rm D+Q}-{\bf d}_{\rm Q}$. We then model this estimator as ${\bf d}_{\rm D}=b_{\rm DLA}{\bf t}_{\rm D}$, where ${\bf t}_{\rm D}$ is a theoretical template for the DLA-only signal with unit bias computed as described in Section \ref{sec:theory}. This method has the advantage that no assumptions are made regarding the particular form of the QSO signal.
        \item {\bf Simultaneous modelling.} We simultaneously model the DLA$+$QSO and QSO-only measurements as
          \begin{align}
            {\bf d}_{\rm D+Q}=b_{\rm DLA}{\bf t}_{\rm D}+b_{\rm QSO}{\bf t}_{\rm Q}+{\bf t}_{\rm M},\hspace{12pt}
            {\bf d}_{\rm Q}=b_{\rm QSO}{\bf t}_{\rm Q}+{\bf t}_{\rm M},
          \end{align}
        where ${\bf t}_{\rm Q}$ is the theoretical template for the QSO signal with unit bias, $b_{\rm QSO}$ is the bias of this sample and ${\bf t}_{\rm M}$ is the contribution from magnification,. This method has the advantage that, by modelling the QSO lensing signal, we reduce the impact of the noise of ${\bf d}_{\rm Q}$ on the final uncertainty on $b_{\rm DLA}$.
      \end{enumerate}
      
      In the following sections we present the measurement of ${\bf d}_{\rm D+Q}$ and ${\bf d}_{\rm Q}$ in both harmonic space and configuration space, as well as the resulting estimates of $b_{\rm DLA}$.
      
    \subsection{Harmonic-space pipeline}\label{ssec:results.harmonic}
      This section describes the process used to estimate the power spectra $C^{\kappa\alpha}_\ell$, with $\alpha$ corresponding to the positions of DLAs or QSOs. The main complication to overcome in this case is the incomplete sky coverage of both the convergence map and the SDSS catalogs, which biases any direct estimate of either field's harmonic coefficients. In order to account for this effect, we use the so-called pseudo-$C_\ell$ (PCL) or MASTER approach \cite{2002ApJ...567....2H}, which we describe briefly here.
      
      In short, we observe incomplete maps of the sky for a couple of fields $f^1$ and $f^2$: $\hat{f}^i(\nv)=w^i(\nv)\,f^i(\nv)$, where $w^i$ is a generic weights map, which can be a binary mask defined by a given experiment's footprint. The harmonic coefficients of $\hat{f}^i$ are therefore a convolution of the harmonic coefficients of $f^i$ and $w^i$: $\hat{f}^i_{\ell m}=\sum_{\ell'm'\ell''m''}D^{mm'm''}_{\ell\ell'\ell''} f^i_{\ell'm'} w^i_{\ell''m''}$. Thus, it is possible to prove that the standard power spectrum estimator in the full sky (essentially averaging over the redundant $m$) yields a biased estimate of the true power spectrum $C^{12}_\ell$:
      \begin{equation}
        \left\langle\hat{C}^{12}_\ell\right\rangle\equiv\frac{1}{2\ell+1}\sum_{m=-\ell}^\ell \left\langle{\rm Re}\left(\hat{f}^1_{\ell m}\hat{f}^{2*}_{\ell m}\right)\right\rangle=\sum_{\ell'} W_{\ell\ell'}C^{12}_\ell,
      \end{equation}
      where $W_{\ell\ell'}$ is a mode-mixing matrix dependent on the properties of the weight maps. The PCL method is based on computing an analytical estimate of $W_{\ell\ell'}$ and inverting the previous equation to remove the bias associated to an incomplete sky coverage. We computed the PCL estimate of the $\kappa$-DLA and $\kappa$-QSO power spectra using the public code {\tt NaMaster}\footnote{\url{https://github.com/damonge/NaMaster/}}. The cross-power spectra were estimated in 20 linearly-spaced bandpowers of width $\Delta\ell=50$ between $\ell=2$ and $\ell=1002$.
      
      The PCL estimator requires a sky map and the associated weights map. Both are directly available in the case of the CMB convergence, as described in Section \ref{sec:data}. For an object catalog, as is the case for the DLA and quasar samples, we must construct our own maps. The weights map should be built from the non-trivial completeness data associated with the DR12 quasar sample, since inaccurately accounting for this completeness introduces artificial inhomogeneities in the associated overdensity map. However, while these inhomogeneities would bias any estimate of the auto-correlation of the DLA/QSO two-point statistics, they are uncorrelated with $\kappa$, and therefore should not introduce any biases in cross-correlation (although they will contribute towards the noise). We verified this by running our pipeline using three different masks for the DR12 data:
      \begin{enumerate}
       \item A high-resolution ($N_{\rm side}=2048$) sky mask built from the random catalogs associated with the BOSS CMASS galaxy sample. Essentially we count the number of random objects in each HEALPix pixel and mask out all empty pixels.
       \item A low-resolution ($N_{\rm side}=64$) mask built from the DR12 QSO catalog, again masking out all pixels with no QSOs in them.
       \item An even lower resolution version of the previous mask with $N_{\rm side}=32$.
      \end{enumerate}
      The final results did not change significantly with the mask used, which demonstrates the robustness of the measurement with respect to this choice.
      
      For both the DLA and QSO samples we build a map of their projected overdensity in each pixel $p$:
      \begin{equation}
        \delta_p\equiv\frac{N_p}{\bar{N}}-1,
      \end{equation}
      where $N_p$ is the weighted number of objects in each pixel, $N_p\equiv\sum_{g\in p}w_g$, and $\bar{N}$ is the average of $N_p$ over all unmasked pixels. As described in Section \ref{sec:results.diffm}, QSOs were weighted to reproduce the properties of the DLA parent catalog, and unit weights were used for the DLAs themselves.
          
    \subsection{Configuration-space pipeline}\label{ssec:results.config}
      We complement the harmonic-space measurement described in the previous section with a configuration-space pipeline based on the angular cross-correlation function $\xi^{\kappa\alpha}(\theta)$. Measurements in real space are arguably more robust in terms of systematics associated with survey geometry definition (e.g. defining the QSO weights map as discussed in the previous section). They also provide a more intuitive observable, since $\xi^{\kappa\alpha}(\theta)$ is essentially the average lensing convergence caused by the DLAs/QSOs as a function of angular separation. In any case, this parallel pipeline allows us to validate the measured value of $b_{\rm DLA}$ in harmonic space.
      
      The estimator we use for $\xi^{\kappa\alpha}$ essentially consists of stacking the $\kappa$ map on the position of the DLAs/QSOs:
      \begin{equation}
        \xi^{\kappa\alpha}(\theta)=\frac{\sum_gw_g\,\bar{\kappa}_g(\theta)}{\sum_gw_g},
      \end{equation}
      where $g$ runs over all objects in either catalog and $w_g$ is the weight associated with the $g$-th object. The mean convergence for each object $\bar{\kappa}$ is estimated as:
      \begin{equation}
        \bar{\kappa}_g(\theta)=\frac{\sum_pw^\kappa_p\kappa_p}{\sum_pw^\kappa_p},
      \end{equation}
      where the weights $w^\kappa_p$ correspond to the mask associated with the convergence map provided by Planck, and $p$ runs over all pixels separated from the position of the $g$-th object by an angle lying in the interval $[\theta-\Delta\theta/2,\theta+\Delta\theta/2]$ (here $\Delta\theta$ is the choice of bin width for our correlation function estimator). For our configuration-space measurement we estimated the correlation function in 16 bins of $\theta$ in the interval $0<\theta<3^\circ$ (i.e. $\Delta\theta\sim11'$).
      
      We pre-process the convergence map by Wiener-filtering it before computing the correlation function. This effectively down-weights noise-dominated modes optimally, reducing their impact in the final $S/N$. The Wiener filter applied to the harmonic coefficients $\kappa_{\ell m}$ is $W_\ell\equiv C^{\kappa\kappa}_\ell/(C^{\kappa\kappa}_\ell+N^{\kappa\kappa}_\ell)$, where $C^{\kappa\kappa}_\ell$ is the convergence power spectrum for the best-fit Planck cosmological parameters, and $N^{\kappa\kappa}_\ell$ is the noise power spectrum of the Planck $\kappa$ map. As mentioned in Section \ref{sec:theory} this weighting is taken into account when modelling the theory prediction for the correlation function. We find that this procedure does not lead to any significant improvement in the results when comparing our pipeline to one that ignores Wiener-filtering, beyond producing more visually-compelling measurements of the lensing signal in real space.
            
    \subsection{Measurement of $b_{\rm DLA}$}\label{ssec:results.measurement}
      As described in Section \ref{sec:results.diffm}, both of our approaches to measure $b_{\rm DLA}$ model the data vector ${\bf d}$ as ${\bf d}=\hat{\sf T}\,{\bf b}$, where ${\sf T}$ is a matrix of theoretical templates and ${\bf b}$ is a vector containing all bias parameters. In the case of direct subtraction ${\bf d}\equiv{\bf d}_{\rm D}$, $\hat{\sf T}\equiv{\bf t}_{\rm D}$ and ${\bf b}=b_{\rm DLA}$, whereas for simultaneous modelling:
      \begin{equation}
        {\bf d}\equiv\left(
        \begin{array}{c}
         {\bf d}_{\rm D+Q}-{\bf t}_{\rm M}\\
         {\bf d}_{\rm Q}-{\bf t}_{\rm M}
        \end{array}\right),\hspace{12pt}
        \hat{\sf T}\equiv\left(
        \begin{array}{cc}
         {\bf t}_{\rm D} & {\bf t}_{\rm Q}\\
         0               & {\bf t}_{\rm Q}
        \end{array}\right),\hspace{12pt}
        {\bf b}\equiv\left(
        \begin{array}{c}
         b_{\rm DLA} \\ b_{\rm QSO}
        \end{array}\right).
      \end{equation}     
      \begin{figure}
        \centering
        \includegraphics[width=0.49\textwidth]{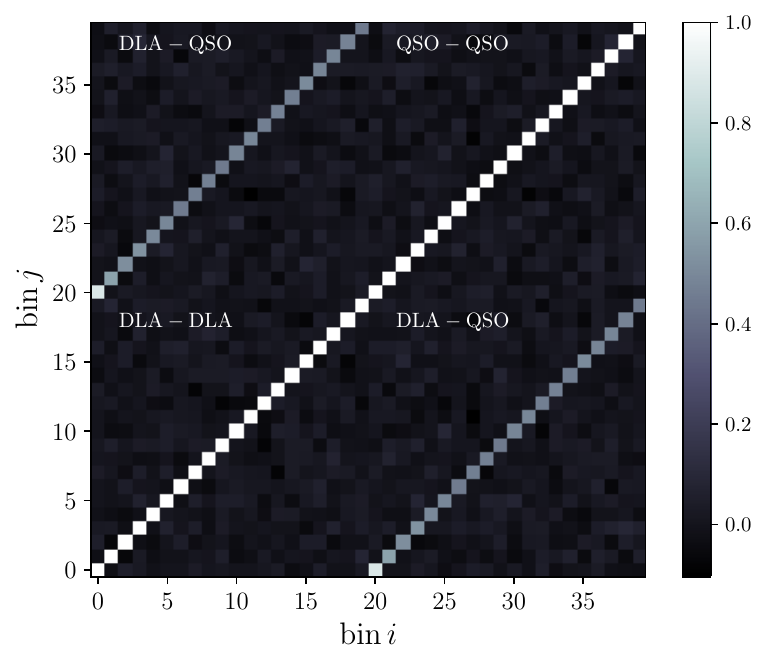}
        \includegraphics[width=0.49\textwidth]{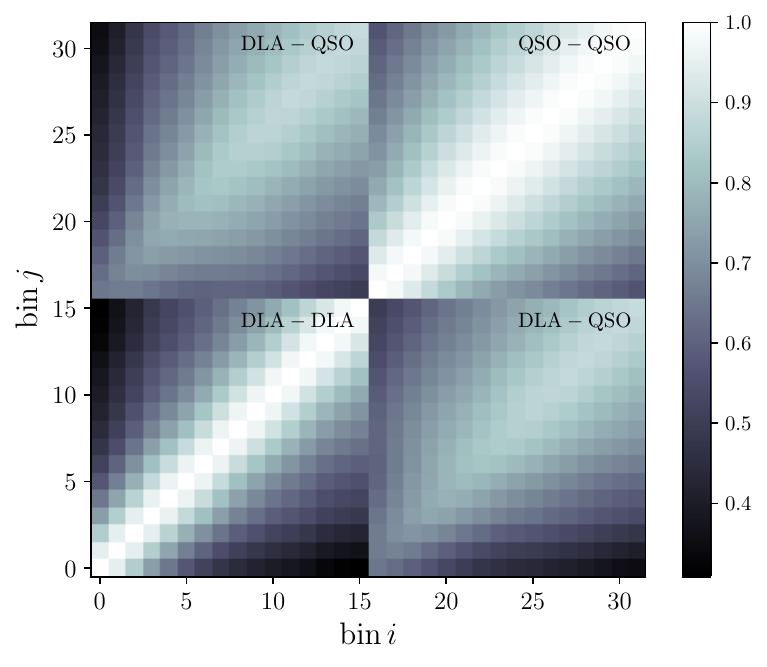}
        \includegraphics[width=0.49\textwidth]{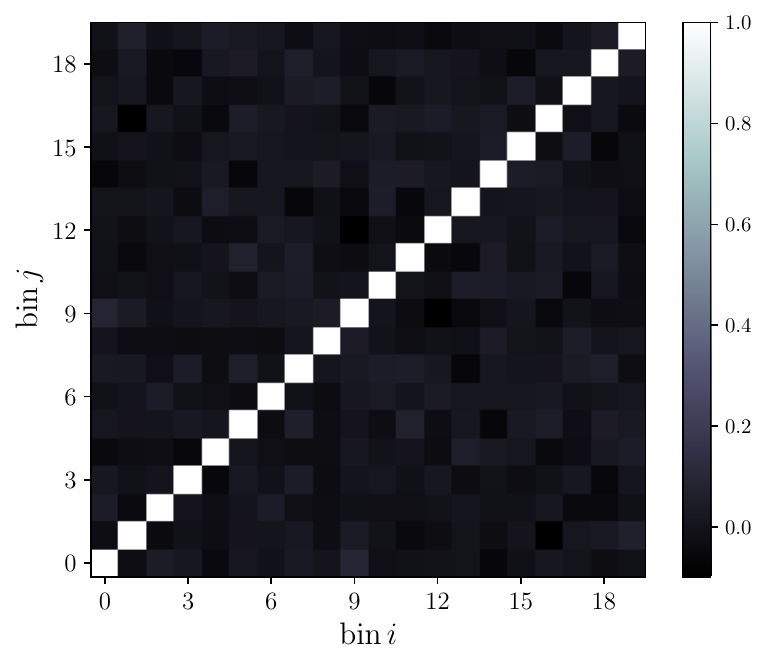}
        \includegraphics[width=0.49\textwidth]{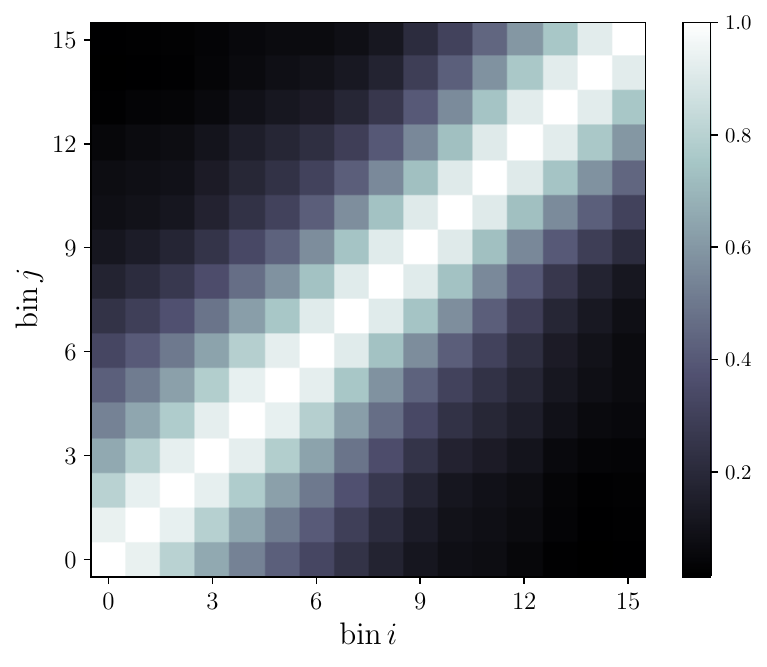}
        \caption{{\sl Top:} data correlation matrix for the simultaneous-fitting pipeline. The four boxes in each plot correspond to the different combinations of pairs of correlations involving DLAs or QSOs (here ``DLA'' denotes the measurement of the lensing signal around the positions of DLAs, which contains the contribution from the background QSO). {\sl Bottom:} data correlation matrix for the direct-subtraction pipeline. {\sl Left:} results the $C_\ell$-based analysis. {\sl Right:} data correlation matrix for the $\xi(\theta)$-based analysis. The large-scale correlations are caused by the strong smoothing associated to the Wiener filter applied to the $\kappa$ map before computing the correlation function.}\label{fig:covar}
      \end{figure}

      Under the assumption that the data ${\bf d}$ is well described by Gaussian statistics, the maximum-likelihood estimate of ${\bf b}$ and its associated covariance are analytically given by
      \begin{equation}\label{eq:estimator}
        \bar{\bf b}=\left(\hat{\sf T}^T\,\hat{\sf C}^{-1}\,\hat{\sf T}\right)^{-1}\,\hat{\sf T}^T\,{\sf C}^{-1}\,{\bf d},\hspace{12pt}
        {\rm Cov}\left(\bar{\bf b}\right)=\left(\hat{\sf T}^T\,\hat{\sf C}^{-1}\,\hat{\sf T}\right)^{-1},
      \end{equation}
      where $\hat{\sf C}$ is the covariance matrix of the data $\hat{\sf C}\equiv\left\langle({\bf d}-\langle{\bf d}\rangle)\,({\bf d}-\langle{\bf d}\rangle)^T\right\rangle$.
      
      \begin{figure}
        \centering
        \includegraphics[width=0.7\textwidth]{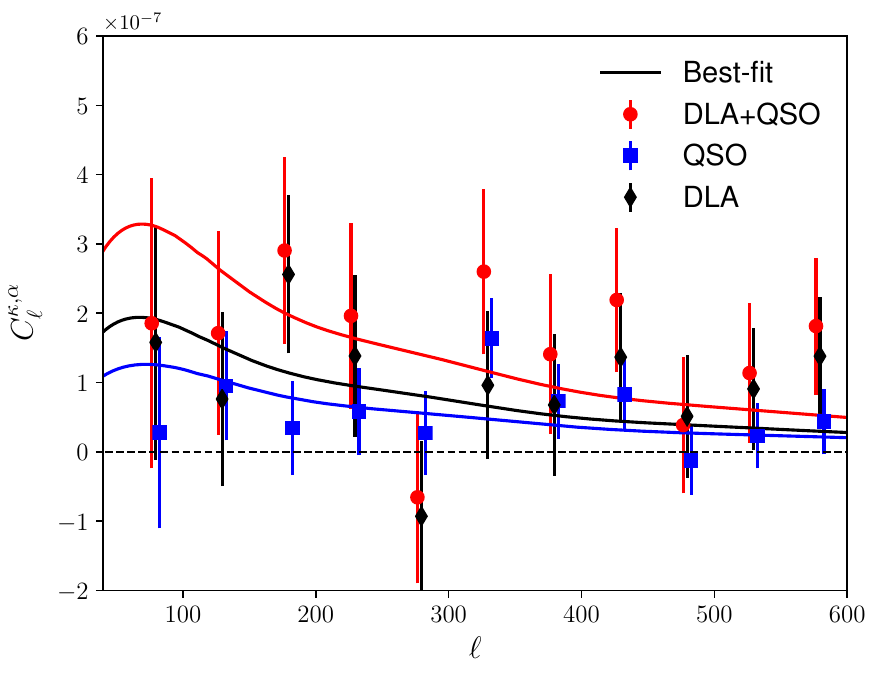}
        \includegraphics[width=0.7\textwidth]{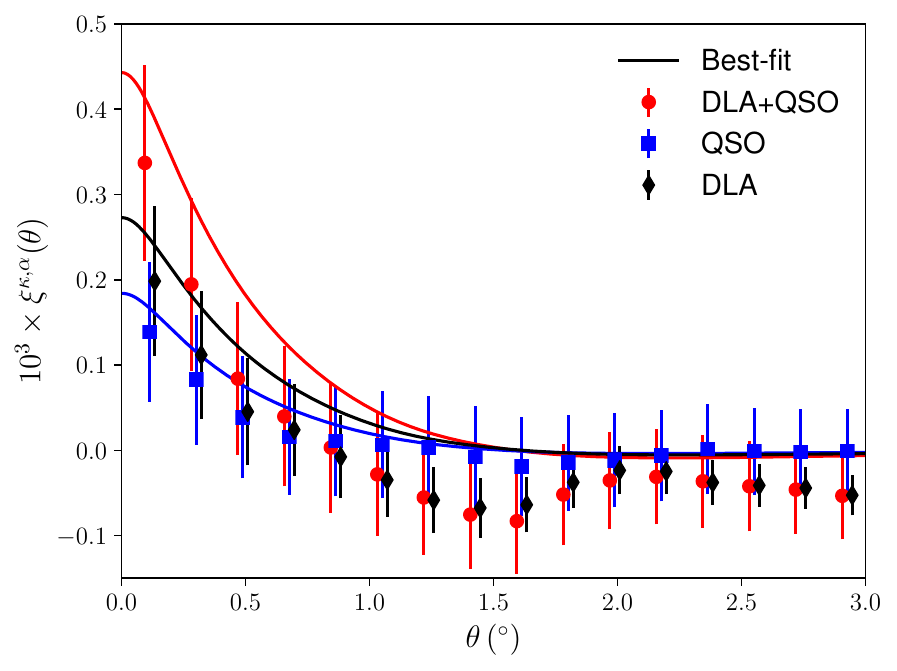}
        \caption{{\sl Top:} measured power spectrum between the CMB lensing convergence $\kappa$ and the raw DLA distribution (red circles), between $\kappa$ and the parent QSO distribution (blue squares) and the difference of both (black diamonds), corresponding to the pure DLA lensing signal with the direct-subtraction pipeline. The solid lines show the best-fit theory predictions. {\sl Bottom:} results for the two-point correlation function with the same color coding. In all cases, the data points are slightly displaced along the $x$ axis to make them distinguishable.}\label{fig:res_fid}
      \end{figure}
\begin{table}
\centering
{
\renewcommand{\arraystretch}{1.2}
\begin{tabular}{c|ccccc}
\hline \hline
DLA catalog & Pipeline & $b_{\rm DLA}\pm1\sigma$ & $b_{\rm QSO}\pm1\sigma$ & $\chi^2/{\rm d.o.f.}$ & PTE\\
\hline
\multirow{4}{*}{N12} 
&$C_\ell$, S.M.      & $2.66\pm0.93$ & $2.90\pm0.69$ & 0.90 & 0.63 \\
&$C_\ell$, D.S.      & $2.55\pm0.94$ &      --       & 0.88 & 0.60 \\
&$\xi(\theta)$, S.M. & $2.56\pm0.91$ & $2.99\pm0.67$ & 0.90 & 0.72 \\
&$\xi(\theta)$, D.S. & $2.69\pm0.92$ &      --       & 1.10 & 0.31 \\
\hline \hline
\multirow{4}{*}{N12B}
&$C_\ell$, S.M.      & $2.01\pm1.01$ & $2.84\pm0.69$ & 1.00 & 0.47 \\
&$C_\ell$, D.S.      & $1.85\pm1.02$ &      --       & 1.10 & 0.32 \\
&$\xi(\theta)$, S.M. & $1.76\pm0.91$ & $2.93\pm0.67$ & 1.08 & 0.31 \\
&$\xi(\theta)$, D.S. & $1.90\pm1.00$ &      --       & 1.30 & 0.11 \\
\hline \hline
\multirow{4}{*}{G16}
&$C_\ell$, S.M.      & $1.92\pm0.69$ & $2.56\pm0.52$ & 1.01 & 0.45 \\
&$C_\ell$, D.S.      & $1.83\pm0.70$ &      --       & 1.02 & 0.43 \\
&$\xi(\theta)$, S.M. & $1.78\pm0.89$ & $2.89\pm0.63$ & 0.80 & 0.87 \\
&$\xi(\theta)$, D.S. & $1.79\pm0.68$ &      --       & 1.33 & 0.10 \\
\hline \hline
\end{tabular}
}    
\caption{Summary of results: best-fit and 1$\sigma$ uncertainties for the bias of DLAs and QSOs, reduced $\chi^2$ and associated probability-to-exceed (PTE). Results are shown for our two different analysis pipelines, direct subtraction (D.S.) and simultaneous modelling (S.M.) in harmonic and real spaces ($C_\ell$ and $\xi(\theta)$ respectly). The three blocks of rows show results for the three different DLA catalogs considered here: N12 (fiducial), N12B and G16.} \vspace{-1.5em}
\label{tab:results}
\end{table}
      We estimated the fiducial data covariance matrix used in our analysis from a sample of 1000 simulated measurements of ${\bf d}$. Each of these were produced by first generating a random realization of the CMB lensing convergence map and correlating it with the positions of DLAs and QSOs in the original catalogs. The random $\kappa$ maps were generated as Gaussian realizations of the best-fit power spectrum of the Planck map, including signal and noise. We verified our estimate of the covariance against the 100 random maps provided by Planck (which better account for the inhomogeneous and non-Gaussian nature of the Planck noise). We also compared the diagonal errors computed this way with analytical ``$1/f_{\rm sky}$'' estimates and with similar random realizations also involving randomly-positioned DLAs and QSOs. Our final results were not found to be sensitive to any of these variations. Figure \ref{fig:covar} shows the covariance matrices used in the different variations of our pipeline (real-space, fourier-space, direct subtraction or joint modelling).
      
      Our measurements of the 2-point statistics and the best-fit theoretical estimates are shown in Figure \ref{fig:res_fid}. The best-fit values of $b_{\rm DLA}$ associated to these measurements are:
      \begin{align}\nonumber
        \text{Direct subtraction:}\hspace{12pt} b_{\rm DLA}=2.55\pm0.94\,\,(C_\ell),\hspace{12pt}b_{\rm DLA}=2.69\pm0.92\,\,(\xi(\theta))\\
        \text{Simultaneous model:}\hspace{12pt} b_{\rm DLA}=2.66\pm0.93\,\,(C_\ell),\hspace{12pt}b_{\rm DLA}=2.55\pm0.91\,\,(\xi(\theta))
      \end{align}
      All our measurements are therefore highly compatible with each other, as well as with the latest measurement of $b_{\rm DLA}=1.99\pm0.11$ from \cite{2018MNRAS.473.3019P}. In all cases the model described in Section \ref{sec:theory} is a good fit to the data, with reasonable $\chi^2$ values and associated probabilities-to-exceed (PTE) above 31\%. Our results (including those discussed in the next section) are listed in Table \ref{tab:results}.
      
      Table \ref{tab:results} shows both our best-fit values for the DLA bias and for the quasar bias. The latter is found to be marginally lower than measurements of the same quantity made using the quasar auto-correlation\footnote{See \cite{2017A&A...608A.130D} for a similar and compatible measurement ($b_{\rm QSO}=3.70\pm0.12$) in cross-correlation with the \lya~forest.} ($b_{\rm QSO}= 3.5\pm0.1$, \cite{2015MNRAS.453.2779E}), and potentially consistent with the results of \cite{2017arXiv170604583D} for a different quasar sample. There are a number of reasons that could explain this slight discrepancy (beyond a simple statistical fluctuation): the different ranges of scales used in each type of  analysis, an incomplete modelling of the clustering properties of quasars, uncertainties in the contribution from cosmic magnification to the cross-correlation of quasars and CMB lensing (e.g. errors in the slope of the quasar luminosity function) or selection effects in the quasar sample. One example of the latter would be the impact of extra-galactic dust, which correlates with the CMB lensing convergence, in the detectability of quasars. Our measurement of the DLA bias is effectively immune to all these sources of systematic uncertainty, since they are subtracted from the DLA$+$QSO measurements together with the quasar lensing signal. One potential caveat to this would be if any of these systematics correlates strongly with the probability for a quasar to host DLAs in its spectrum, although we have not been able to find an instance of this. If we were to fix the QSO bias to the value measured in auto-correlation, $b_{\rm QSO}\simeq3.5$, the best-fit DLA bias drops slightly to $b_{\rm DLA}=2.23\pm0.94$, still in agreement with the measurements from the cross-correlation with the \lya~forest.

    \subsection{Consistency tests}\label{ssec:results.consistency}
      In the previous sections we have shown that our measurement is robust against a number of effects, such as the choice of Fourier-space vs. configuration-space analysis, the modelling of the clustering of background QSOs, the definition of the DLA/QSO mask or the method used to estimate the covariance matrix. In this section we explore other possible systematic effects, especially associated with the properties of our fiducial DLA sample. The values of $b_{\rm DLA}$ quoted in this section will correspond to those measured using a $C_\ell$-based analysis and simultaneously fitting $b_{\rm QSO}$, although we find consistent values with all other analysis methods in all cases.
      \begin{figure}
        \centering
        \includegraphics[width=0.75\textwidth]{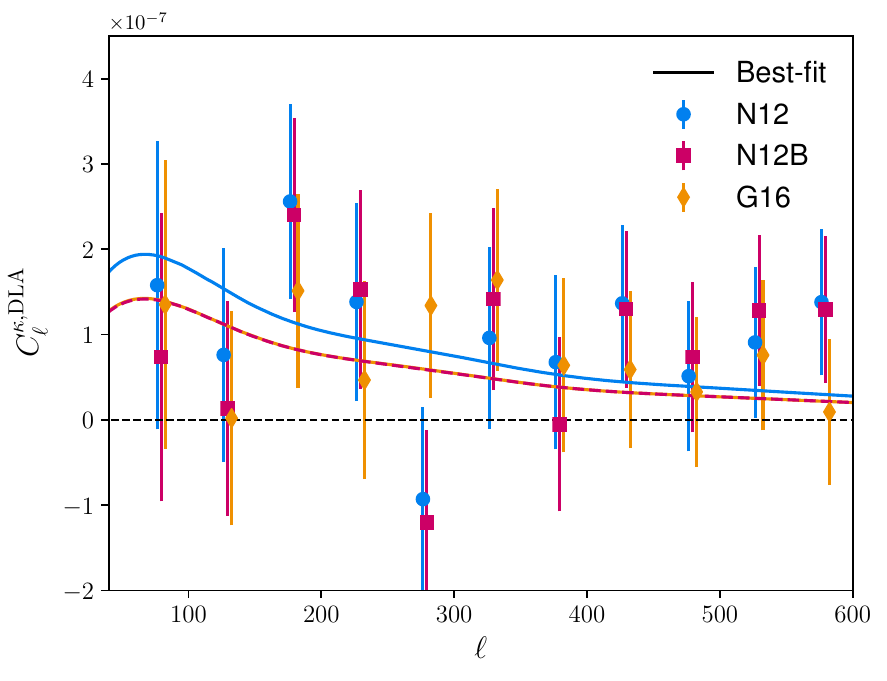}
        \caption{Cross-power spectrum between the CMB lensing convergence and three DLA catalogs: N12 (cyan circles), N12B (dark magenta squares) and G16 (light orange diamonds). In all cases, the contribution from the parent quasar catalog has been directly subtracted. The lines show the best-fit model for each measurement. In all cases, the data points are slightly displaced along the $x$ axis to make them distinguishable.}\label{fig:res_sys}
      \end{figure}
      
      Our fiducial analysis makes use of the raw DLA sample, with no further quality cuts. Given the low signal-to-noise ratio of this measurement, the effects of spurious detections in the sample are likely to be negligible. However, a large number of fake DLAs in the catalog would systematically lower the measurement of $b_{\rm DLA}$ (since they would be uncorrelated with $\kappa$), and therefore it is worth verifying whether such a systematic shift is observed when using a cleaner sample. We do so by repeating our analysis on the so-called N12B catalog, defined by removing all DLAs on spectra with continuum-to-noise ratio ${\rm CNR}<2$, and containing $\sim10\%$ fewer objects. The power spectrum measured on N12B for the direct-subtraction pipeline is shown as dark magenta squares in Fig. \ref{fig:res_sys}, together with our fiducial measurement on N12. We obtain a best-fit measurement of the DLA bias from N12B of $b^{\rm N12B}_{\rm DLA}=2.00\pm1.01$. This corresponds to a $\sim0.7\sigma$ downward shift with respect to our fiducial measurement, with a larger measurement error due to the smaller size of this sample. Since any contamination caused by spurious detections would have produced an upwards shift when removing them, we conclude that our fiducial measurement is robust with respect to this effect.
      
      There exist different procedures to detect DLAs in quasar spectra, and we have also explored the robustness of our results with respect to the definition of DLA used. To do so, we have repeated our analysis on an alternative catalog, G16, produced by \cite{2017MNRAS.472.1850G}. This sample was generated by modelling the quasar spectrum as a Gaussian process, and then determining the detection probability of each DLA using a Bayesian model. We select a clean subsample of this catalog, given by all objects with probability $p>0.9$, containing 35,621 objects. As shown in Section \ref{sec:data}, although the redshift distribution of these DLAs is very similar to that of the N12 sample, the distributions of their parent QSO samples are markedly different (see Fig. \ref{fig:nz}). In practical terms, this implies that the raw DR12 QSO catalog must be re-weighted in order to reproduce the new redshift distribution. This modifies the measured quasar power spectrum and therefore the associated measurement of $b_{\rm DLA}$. The DLA-$\kappa$ power spectrum measured by direct subtraction on G16 is shown as light orange diamonds in Fig. \ref{fig:res_sys}, and we obtain an associated best-fit value $b^{\rm G16}_{\rm DLA}=1.92\pm0.69$.
      
      This measurement is $\sim1.1\sigma$ away from the value found for N12, it has a similar detection significance, and both measurements are compatible with the latest measurement of \cite{2018MNRAS.473.3019P}. However, since both measurements are made on catalogs with a significant overlap, they are not completely uncorrelated. Therefore it is important to assess the degree of disagreement between both measurements more precisely. From the 1000 simulations used to estimate the data covariance matrix we can determine both measurements to be $\sim50\%$ correlated, i.e.:
      \begin{equation}
        r\equiv\frac{{\rm Cov}({\rm N12},{\rm G16})}{\sqrt{{\rm Var}({\rm N12}){\rm Var}({\rm G16})}}=0.501.
      \end{equation}
      Now, let $b_1$ and $b_2$ be two measurements of $b_{\rm DLA}$ with covariance matrix ${\sf C}$. Then, assuming Gaussian statistics, the difference $\Delta b\equiv b_1-b_2$ should have a Gaussian distribution with standard deviation $\sigma(\Delta b)=\sqrt{C_{11}+C_{22}-2C_{12}}$. Thus we can quantify the disagreement between both measurements in terms of the probability of a value $\Delta b$ being larger than our measurement. Doing this, we obtain a probability ${\rm PTE}=0.19$, and therefore the measurements made in N12 and G16 are compatible.

      Finally, we have also explored the impact of other variables in our analysis, including our choice of mask apodization in the $C_\ell$-based pipeline and the binning scheme used both in $\theta$ and $\ell$. None of these factors significantly impact our results. Our confidence in the robustness of our analysis pipeline is further strengthened by our ability to reproduce the measurement of $b_{\rm QSO}$ presented in \cite{2017arXiv170604583D}.

  \section{Forecasts}\label{sec:forecast}
    Although the measurement of $b_{\rm DLA}$ presented here serves to validate previous estimates of this quantity made in cross-correlation with the \lya\ forest $b_{\rm DLA}\sim2$, the noise in the CMB lensing map, as well as the low density of detected DLAs prevent us from discriminating between the measured DLA bias and the lower value $b_{\rm DLA} \sim 1.2$ that is expected from hydrodynamical simulations without strong galactic wings (see Figure 14 in \cite{2012JCAP...11..059F}). It is therefore relevant to explore the potential of future quasar surveys and CMB experiments to improve on this result. To do so, we have estimated the uncertainty $\sigma(b_{\rm DLA})$ for a number of future facilities using a Fisher-matrix formalism (see e.g. \cite{1998ApJ...499..555T}). We have explored three different quasar samples:
    \begin{enumerate}
     \item The SDSS DR12 sample described in Section \ref{sec:data}.
     \item The sample achievable by the Dark Energy Spectroscopic Instrument (DESI, \cite{2013arXiv1308.0847L}). We define this as having a 3-times larger density of QSOs and DLAs covering 14,000 deg$^2$ of the northern celestial hemisphere.
     \item A southern-hemisphere version of the DESI sample, which could be obtained by the 4MOST instrument \cite{2016SPIE.9908E..1OD}. We consider this sample in order to explore the benefit of having a full area overlap with southern-hemisphere, ground-based CMB experiments.
    \end{enumerate}
    \begin{figure}
      \centering
      \includegraphics[width=0.75\textwidth]{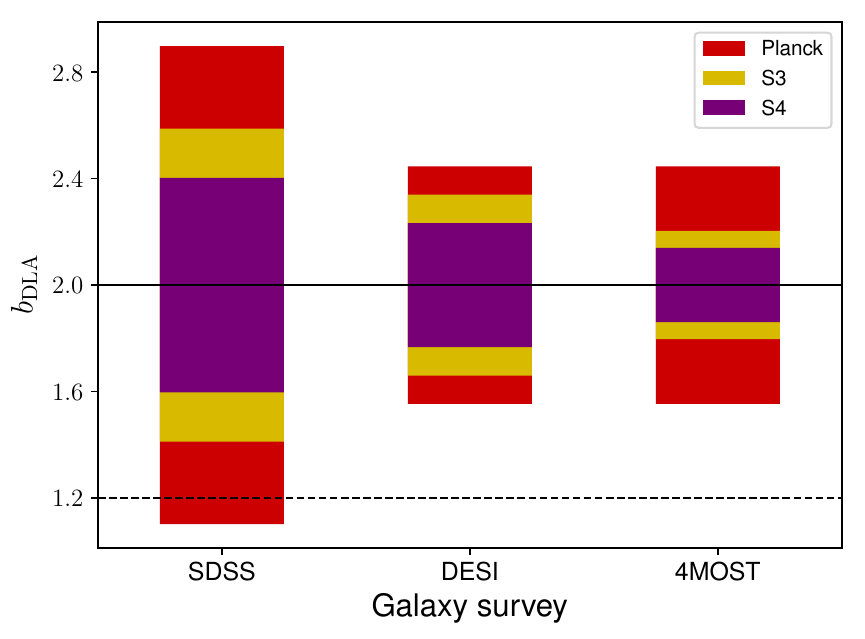}
      \caption{Forecast $1\sigma$ uncertainties on $b_{\rm DLA}$ achievable by current and future surveys and CMB experiments. The horizontal lines show the preferred values for this quantity measured in cross correlation with the \lya\ forest (solid) and estimated from hydrodynamical simulations (dashed). Although current data is unable to discriminate between both cases, a combined analysis of DESI and Stage-4 CMB would make this possible.}\label{fig:4casts}
    \end{figure}
    We studied the constraints on $b_{\rm DLA}$ achievable by each of these surveys in combination with a CMB lensing map produced by three different experiments:
    \begin{enumerate}
      \item The Planck satellite (as described in Section \ref{sec:data}). Given the full-sky coverage provided by Planck, we assume full area overlap with the three previous samples.
      \item A Stage-3 (S3) ground-based CMB experiment, such as Advanced ACT \cite{2016SPIE.9910E..14D}. We characterize this experiment by a 1.4 arcmin FWHM Gaussian beam and a white noise level of $8.5\mu{\rm K}\,{\rm arcmin}$ spread over an area of 20,000 ${\rm deg}^2$ in the southern hemisphere.
      \item A Stage-4 (S4) ground-based experiment \cite{2016arXiv161002743A}, also observing 20,000 ${\rm deg}^2$ of the southern hemisphere, with a noise level of $1\mu{\rm K}\,{\rm arcmin}$ and a 3 arcmin beam.
    \end{enumerate}
    We assumed a total area overlap of only 4,000 ${\rm deg}^2$ between SDSS/DESI and S3/S4, and a full overlap (14,000 ${\rm deg}^2$) with 4MOST.
    
    The forecast values of $\sigma(b_{\rm DLA})$ for the different combinations of quasar sample and CMB experiment are shown in Figure \ref{fig:4casts}. The forecast for SDSS $\times$ Planck agrees well with the errors obtained in this work, and shows that current sensitivities and number densities are insufficient to discriminate between $b_{\rm DLA}=2$ and $b_{\rm DLA}=1.2$. However, the combination of DESI and S4 would achieve $\sim10\%$ errors on this parameter, making the distinction between these two models possible at the $\gtrsim3\sigma$ level. Optimal results would be obtained by maximizing area overlap, which a survey like 4MOST could achieve.

  \section{Conclusions}\label{sec:conclusions}
    In this paper we have measured the bias of DLAs from the cross-correlation between DLAs found in \lya\ forest spectra of quasars and the weak lensing map obtained from the Planck satellite. This is the first measurement of DLA properties based on cross-correlation with weak-lensing. We measure the DLA bias to be $b_{\rm DLA}=2.6\pm0.9$. The formal statistical significance is at 2-3$\sigma$ level, but since DLA clustering has been measured before using cross-correlations with other tracers, the correct way to look at our measurement is as a noisy measurement of the bias of DLAs. We have performed the measurement in both Fourier and configuration space, and performed a number of tests and data-splits. We find our result to be robust against a number of analysis choices. Measurements of the DLA bias based on cross-correlations with lensing fields do not rely on the bias model of any other tracer and derive their information on the very large scales where theoretical modeling is least prone to systematic uncertainties. 

    Our measurement is consistent with the best measurements derived from the cross-correlation between the \lya\ forest fluctuations and the DLA density field ($b_{\rm DLA}=1.99\pm0.11$), but does not add to the discussion of this surprisingly high value given our current understanding of astrophysics relevant to DLAs, since our result is also consistent with the considerably lower values favored by hydro simulations $b_{\rm DLA}\sim 1.2$. 

    We forecast the ability of future measurements to improve on this number and find that, with DESI or 4MOST in combination with future Stage-3 and Stage-4 CMB experiments, we will have statistical power to decisively differentiate between the high-bias scenario currently favored by data and low-bias scenario favored by simulations. With improvement in the lensing field, these measurements will also become statistically competitive with forest cross-correlation measurements. Resolving the DLA bias puzzle is not important just for the astrophysics of low-mass halos, but it also have repercussions for predicting future 21cm cosmology signals.

\acknowledgments

We would like to thank Simeon Bird, Emanuele Castorina, Jordi Miralda-Escud\'e, Pasquier Noterdaeme, J. Xavier Prochaska and Francisco Villaescusa-Navarro for useful comments and discussions. We also acknowledge the anonymous journal referee for their comments, which helped us improve the quality of this paper. DA thanks the Center for Computational Astrophysics, part of the Flatiron Institute, for their hospitality during part of this work. DA acknowledges support  from  the  Science  and  Technology  Facilities  Council (STFC) and the Leverhulme and Beecroft Trusts. AFR acknowledges support by an STFC Ernest Rutherford Fellowship, grant reference ST/N003853/1. AS acknowledges hospitality of the Cosmoparticle Hub at the University College London during which parts of this work have been completed. This work was partially enabled by funding from the UCL Cosmoparticle Initiative.

\bibliographystyle{JHEP}
\bibliography{main}

\end{document}